\documentclass[useAMS,usenatbib]{mn2e}
\usepackage{graphicx}
\def\cbeta{$c_{\beta}$}  
\def\kms{\relax \ifmmode {\,\rm km\,s}^{-1}\else \,km\,s$^{-1}$\fi}

\def\mincir{\ \raise-2.truept\hbox{\rlap{\hbox{$\sim$}}\raise5.truept
    \hbox{$<$}\ }}
\def\magcir{\ \raise-2.truept\hbox{\rlap{\hbox{$\sim$}}\raise5.truept
    \hbox{$>$}\ }}

\def\arcsec{\hbox{$^{\prime\prime}$}}

\def\niii{N {\sc iii}}
\def\hi{H~{\sc i}}
\def\hii{H~{\sc ii}}
\def\sii{[S~{\sc ii}]}

\def\oii{[O~{\sc~ii}]}

\def\heii{He~{\sc ii}}
\def\hei{He~{\sc i}}
\def\oiii{[O~{\sc iii}]}
\def\ovi{O~{\sc vi}}
\def\neiii{[Ne~{\sc iii}]}
\def\nevii{Ne~{\sc vii}}
\def\fevii{[Fe~{\sc vii}]}
\def\fev{[Fe~{\sc v}]}
\def\nev{[Ne~{\sc v}]}

\def\ha{H$\alpha$}
\def\hb{H$\beta$}
\def\hd{H$\delta$}   
\def\hg{H$\gamma$}   

\def\te{$T_e$}

\def\teff{$T_{eff}$}

\def\n205{NGC~205}
\def\dI{dIrr}
\def\dS{dSph}

\title[Discovery of symbiotic stars in NGC~205]{{Discovery of true, likely and possible symbiotic stars 
in the dwarf spheroidal NGC~205}\thanks{Based on observations obtained at the Gemini 
Observatory, which is operated by the Association of Universities for Research in Astronomy, 
Inc., under a cooperative agreement with the NSF on behalf of the Gemini partnership.}}

\author[D. R. Gon\c calves et al.]{Denise R. Gon\c calves$^{1}$\thanks{E-mail:
denise@astro.ufrj.br}, Laura Magrini$^{2}$, Ignacio G. de la Rosa$^{3,4}$, and Stavros Akras$^{1}$
\\
  $^{1}$ Observat\'orio do Valongo, Universidade Federal do Rio de Janeiro, Ladeira Pedro Antonio 43, 20080-090 Rio de Janeiro, Brazil\\
  $^{2}$ INAF - Osservatorio Astrofisico di Arcetri, Largo E. Fermi 5, I-50125 Firenze, Italy\\
  $^{3}$ Instituto de Astrof\'\i sica de Canarias, E-38205 La Laguna, Tenerife, Spain\\
  $^{4}$ Universidad de La Laguna, Departamento de Astrof\'\i sica, E-38206 La Laguna, Tenerife, Spain\\
}

\begin{document}

\date{Accepted ?. Received ?; in original form ?}

\pagerange{\pageref{firstpage}--\pageref{lastpage}} \pubyear{2013}

\maketitle

\label{firstpage}

\begin{abstract}
{In this paper we discuss the photometric and spectroscopic observations of newly discovered (symbiotic) systems in the dwarf spheroidal galaxy NGC~205. The Gemini Multi-Object Spectrograph on-off band \oiii\ 5007\AA\ emission imaging highlighted several \oiii\ line emitters, for which optical spectra were then obtained (Gon\c calves et al. 2014). The detailed study of the spectra of three objects allow us to identify them as true, likely and possible symbiotic systems (SySts), the first ones discovered in this galaxy. SySt-1 
is unambiguously classified as a symbiotic star, because of the presence of unique emission lines which belong \textit{only} to symbiotic spectra, the well known \ovi\ Raman scattered lines. SySt-2 is only possibly a SySt because the Ne~{\sc vii} Raman scattered line at 4881\AA, recently identified in a well studied Galactic symbiotic as another very conspicuous property of symbiotic, could as well be identified as {\sc N~III} or [Fe~{\sc III]}.  
Finally, SySt-3 is likely a symbiotic binary because in the red part of the spectrum it shows the continuum of a late giant, and forbidden lines of moderate- to high-ionization, like \fev\ 4180\AA. The main source for skepticism on the symbiotic nature of the latter systems is their location in the PN region in the \oiii4363/\hg\ vs \oiii5007/\hb\ diagnostic diagram (Gutierrez-Moreno et al. 1995). It is worth mentioning that at least another two confirmed symbiotics, one of the Local Group dwarf spheroidal IC~10 and the other of the Galaxy, are also misplaced in this diagram.}
\end{abstract}


\section[]{Introduction}
Symbiotic stars (SySts) are binary systems composed by a cool giant star and a hot companion, which 
can be a white dwarf, a main sequence star with accretion disc, or a neutron star. 
The wind expelled by the cool giant star accretes onto the hot companion powering the 
symbiotic activity, including occasional eruptions and jets. 

Symbiotic systems display characteristic spectra, which allow their detection with on-off band techniques, 
in which narrow-band filters are tuned in to one or more typical emission lines. In an attempt to separate symbiotic systems from their most common mimic Magrini, Corradi \& Munari (2003) used several diagnostic diagrams based on the fluxes obtained with the narrow-band filters H$\alpha$ and \oiii\ at 5007\AA. They concluded that it is, indeed, possible  to discover symbiotic systems using their proposed diagrams, but it is not easy to discriminate them from planetary nebulae (PNe) or compact \hii\  regions without a spectroscopic follow-up. They also estimated  the  expected number of symbiotic stars in a given galaxy, finding that it increases with the luminosity (mass) of the galaxy. For a galaxy like \n205, the second most massive dwarf companion of Andromeda (3.3$\times$10$^8$M$_{\odot}$; McConnachie 2012), the predicted number of symbiotic stars is 1.7$\times$10$^4$ (Magrini et al.  2003), including both active and inactive systems. Active systems are spectroscopically detectable because their stars are interacting and exchanging matter.

This figure is an absolute estimation, and, obviously, it has to be rescaled to obtain the number of SySts that can actually be observed. If the ratio of known (up to 300, Miszalski et al. 2014; Rodr\'iguez-Flores et al. 2014) versus expected ($~$400,000, Magrini et al.  2003) SySts in our Galaxy is assumed to be valid in \n205, around $\sim$12 SySts are expected to be active and detectable in \n205. In general, SySts are expected to be more abundant in dwarf spheroidal (\dS) than in dwarf irregular (\dI) galaxies (Magrini et al. 2003). 

In the present work, the discovery of the first three (true, likely and possible) SySts of \n205 is reported as a result of our spectroscopic study of the emission-line population of the galaxy. The paper is organized as follows: in Sec.2 we describe our observations, while in Sec.3 we present the spectroscopic characterization of the three systems. In Sec.4 we give an estimate of the stellar parameters of the hot and cool companions in the objects. The location of the newly discovered systems in the Gutierrez-Moreno et al. (1995) diagnostic diagram, aimed to separate SySts from PNe, is discussed in Sec.5. We summarize our work and give our conclusions in Sect. 6. 

\section[]{Observational Data}

\begin{figure} 
   \centering 
  \includegraphics[width=8.5truecm]{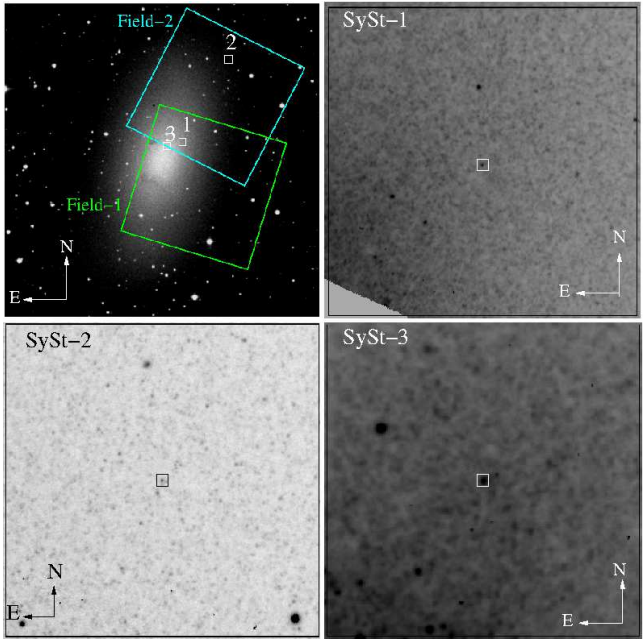}
  \caption{{\it Top right}: A 14.2$\times$14.2~arcmin$^2$ POSSII image of NGC~205, retrieved 
  from the NASA/IPAC Extragalactic Database. The two FoV (5.5$\times$5.5~arcmin$^2$), observed 
  with GMOS@Gemini, are superimposed on the POSSII image. The position of the three symbiotic 
  systems are also highlighted with small boxes within these two FoVs. 1.0$\times$1.0~arcmin$^2$ finding charts of the 
  newly discovered systems: 
  {\it Top right}: centred on SySt-1. {\it Bottom left}: centred on
    SySt--2. {\it Bottom right}: centred on SySt-3. 
   }
   \label{fchart}
\end{figure}

\begin{table}
\centering
\begin{minipage}{75mm}
{\scriptsize  
\caption{Coordinates of the three SySts. F1 (F2) stands for Field-1 (Field-2) GMOS@Gemini FoV we observed. See Figure~1.}
\begin{tabular}{@{}lllcc@{}}
\hline
Field-ID & Name           &  RA          &Dec           & C05$^{\alpha}$  \\
         &                &  J2000.0     &J2000.0       &      \\  	    
\hline
F2-9     & NGC~205 SySt--1 &  00:40:17.60 & +41:41:53.30  & -    \\
F2-8     & NGC~205 SySt--2 &  00:40:07.97 & +41:45:23.64  & PN18 \\
F1-18    & NGC~205 SySt--3 &  00:40:20.97 & +41:41:42.60  & PN36 \\
\hline					  
\multicolumn{5}{l}{$^{\alpha}$ PN ID used by Corradi et al. (2005, C05).} \\
\end{tabular}
}
\end{minipage}
\label{tab_objid}
\end{table}

We obtained the present data by using the imager and spectrograph GMOS, Gemini Multi-Object Spectrograph, at the GEMINI North telescope, in 2010 (Program GN-2010B-Q-107). The GMOS has a 5.5$\times$5.5~arcmin$^2$ field of view, which was centred at the RA/DEC 00:40:12.50/+41:40:03 (Field-1; F1) and 00:40:10.50/+41:43:47.0 (Field-2; F2) of NGC~205, as indicated in  Fig.~1.

We adopted the on-off band imaging technique to identify strong emission-line objects (mainly PNe) with the 
on-band filter \oiii, OIII\_G0318, and the off-band one, \oiii-Continuum, being OIII\_G0319. The central wavelength and width of these two filters are 499~nm/5~nm, and 514~nm/10~nm, respectively. Three exposures of 540(810)~seconds using the on-band(off-band) filter were taken, in July 9 (F1) and September 2-3 (F2) of 2010. We then used the two combined narrow-band frames to build  an \oiii\ continuum-subtracted image of the field, from which we identified a number of objects to be spectroscopically observed, including previously known PNe and other \oiii\ emission-line objects. In Table~1 we also give the Field-ID of the 
symbiotics, for the sake of consistence with the results for the PNe, which were discussed in a previous 
paper (Gon\c calves et al. 2014, hereafter G14).

The spectroscopy of F1 and F2 was obtained with two gratings, R400+G5305 and B600, from 8th to 11th of October 2010. The effective spectral coverage of the spectra is 3600--9400~$\pm$~400~\AA, with initial and final $\lambda$ depending on the location of the slit. Exposure times were of 3$\times$2,400~sec per grating. The disperser central wavelength was slightly varied from exposure to exposure (750 $\pm 10$~nm for R400+G5305, and B600 was centred at 460 $\pm 10$~nm). The slit width was 1\arcsec,  while the slit height was 5\farcs6. The final (binned) spatial scale and reciprocal dispersions of the spectra are as follows: 0\farcs161 and 0.09~nm per 
pixel, with B600; and 0\farcs161 and 0.134~nm per pixel, with R400+G5305. The seeing varied from $\sim$0\farcs42 to $\sim$0\farcs57. CuAr lamp exposures were obtained with both gratings, in the day before or after the science exposures, for wavelength calibration. On October 8th and 12th, 2010 we obtained spectra of spectrophotometric standards (Massey et al. 1988; Massey \& Gronwall 1990)
with the same instrumental setups as for science exposures and we used them to flux calibrate the spectra. The whole reduction and calibration was performed in the standard way by using the Gemini {\sc gmos data reduction script} and {\sc long-slit} tasks, both being part of {\sc IRAF}. We refer the reader to G14 for further details on the observations and reduction where we described, among others details, the effects of the differential atmospheric refraction on the present data. In G14 we also show the good agreement between 
our measured spectroscopic fluxes and those extracted from the literature for common objects.

\subsection{Spectroscopic Measurements and Analysis}

\begin{figure} 
   \centering
   \includegraphics[width=8.25truecm]{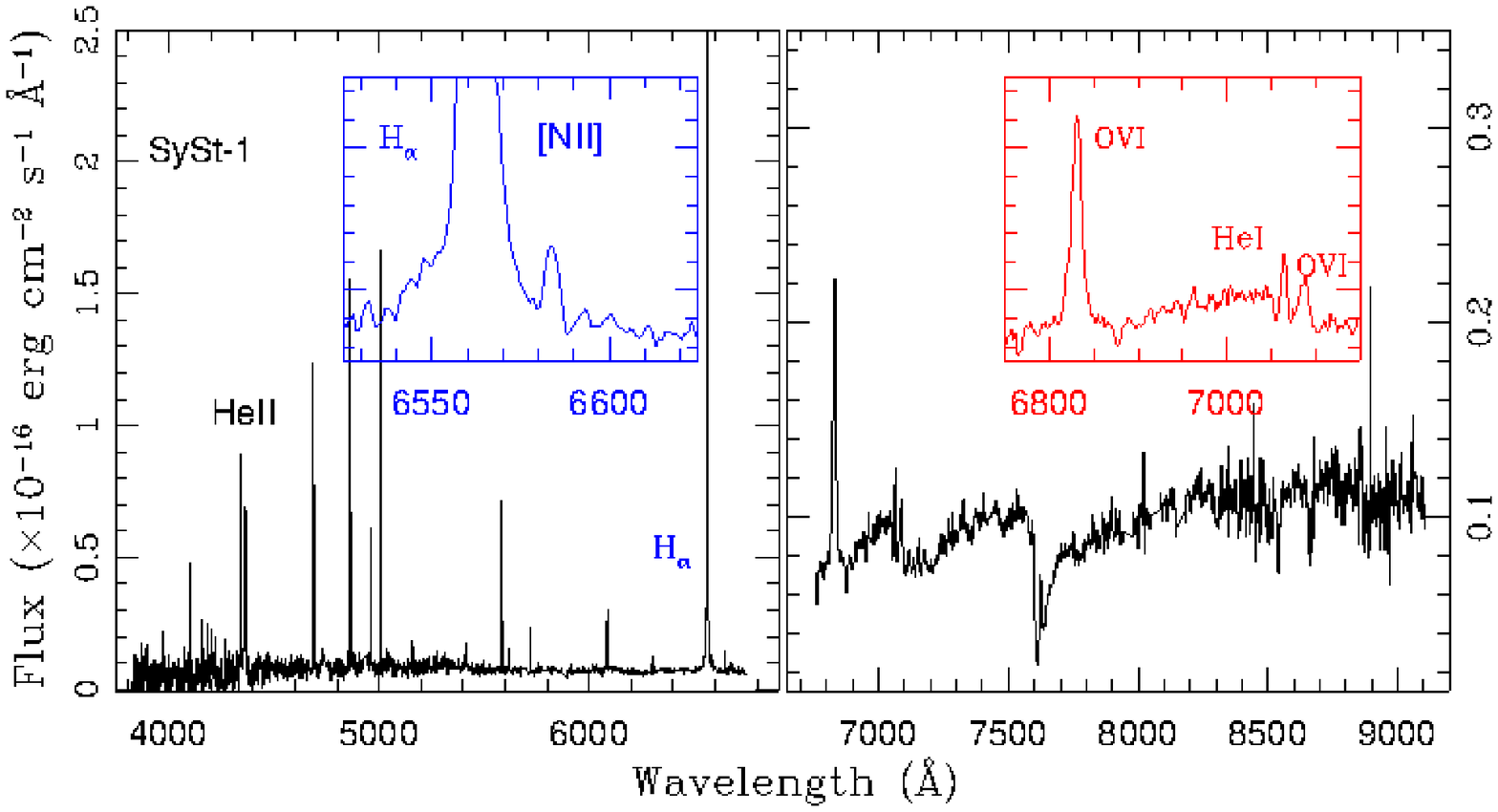}
   \includegraphics[width=8.25truecm]{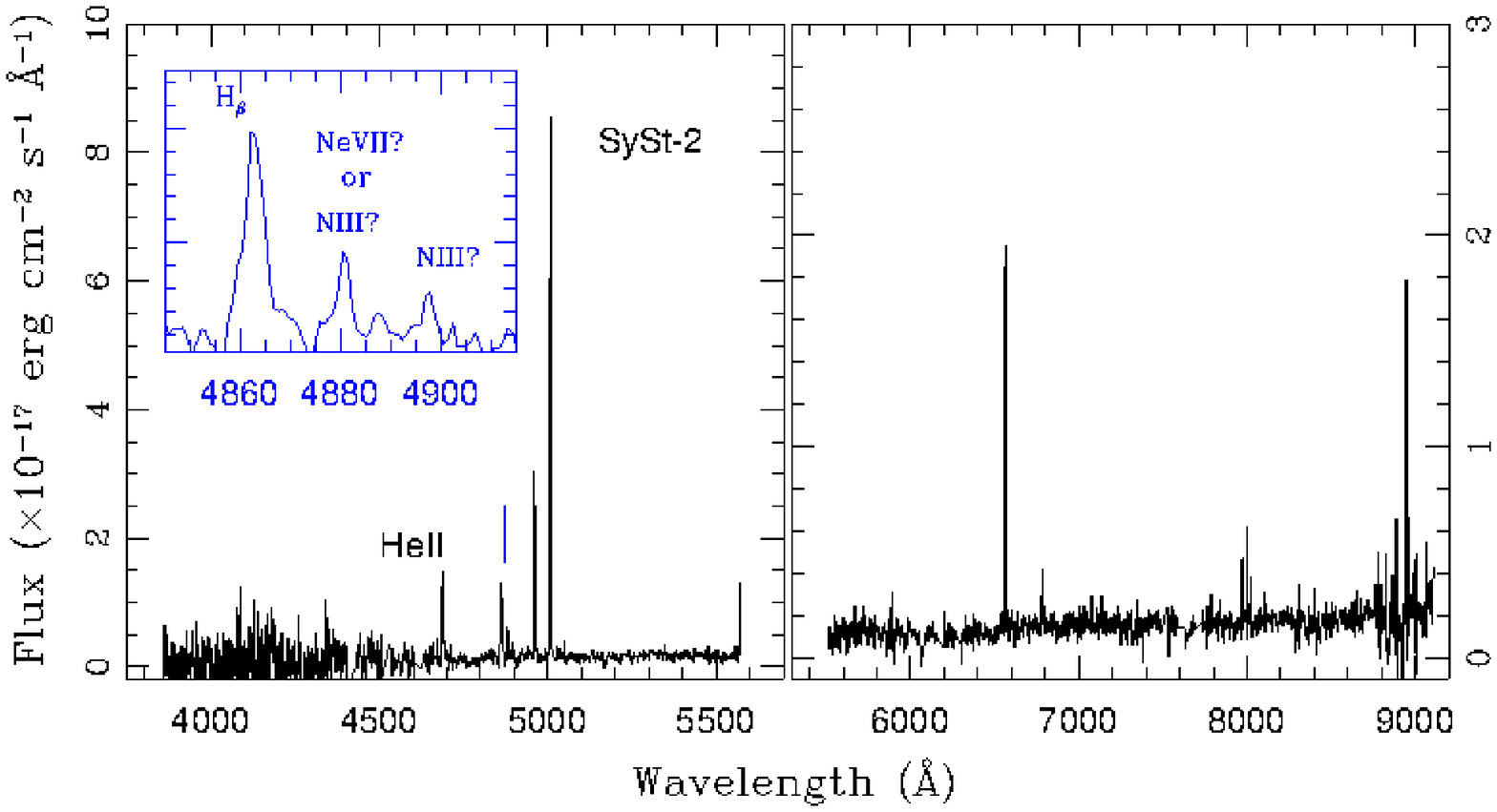}
   \includegraphics[width=8.25truecm]{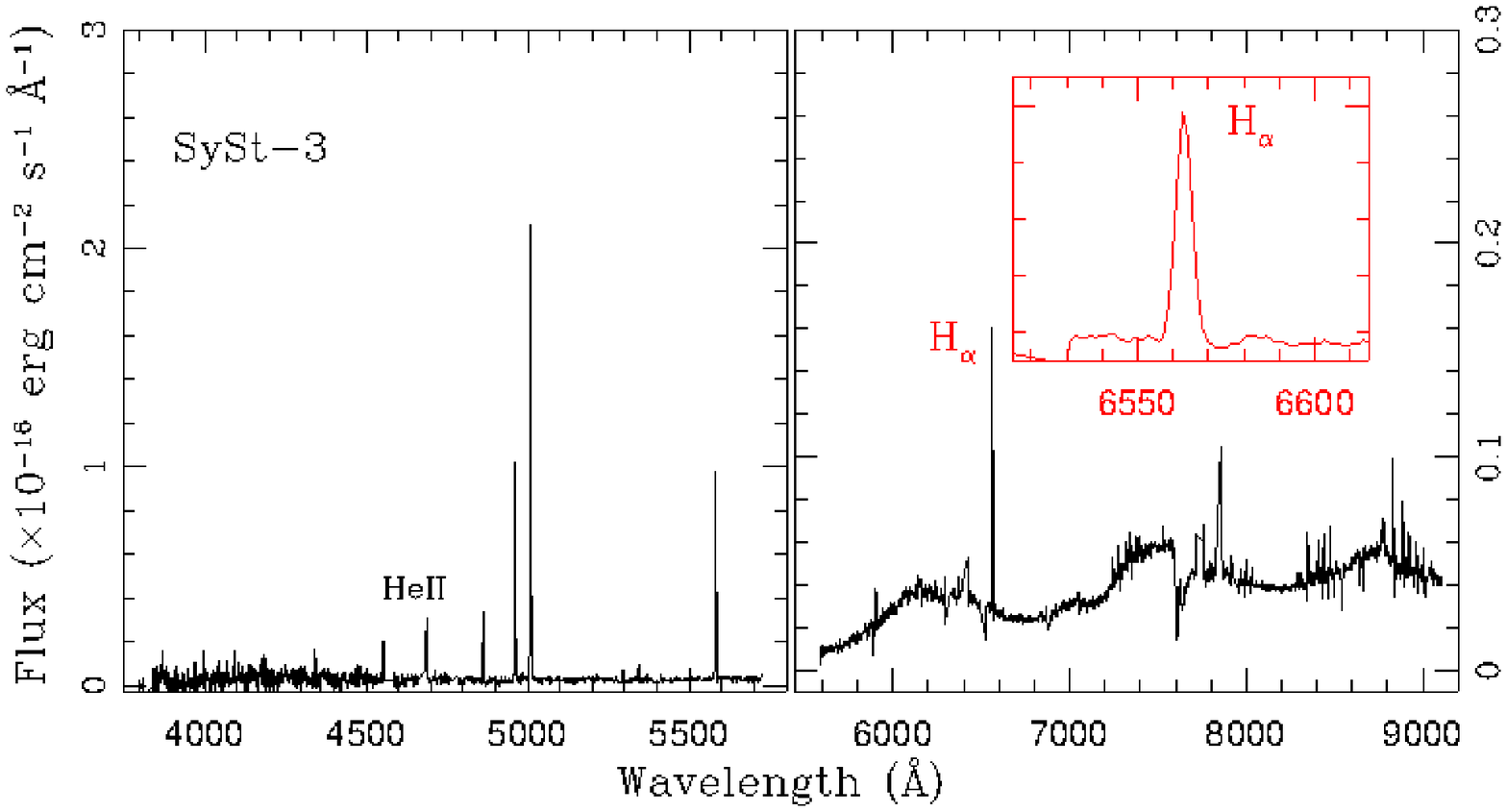}
  \caption{GMOS blue plus red spectra of \n205 symbiotic systems.  
  {\it SySt--1}: 
  The inset at the blue part of the spectrum shows the typical single-peaked broad \ha\ 
  profile, with a shoulder at the blue side, similar to well-known Galactic 
  symbiotics (Tomova \& Tomov 1999; Leedj\"arv et al. 2004). The red side inset highlights 
  the \ovi\ Raman scattered lines (at 6830\AA\ and 7088\AA), which are also broad if compared to 
  the forbidden   lines present in the spectrum (Schmid et al. 1999). Note the intense red continuum of 
  the companion star that makes the fluxes measured in this part of the spectrum very uncertain. 
  {\it SySt--2}: 
  The inset shows the \hb\ as well as the  Raman-scattered Ne~{\sc vii} $\lambda$973 (see Lee, Heo \& 
  Lee~2014) or \niii\ at 4881\AA\ line. Another, not well-identified, line also shows up as a relatively 
  intense line at 4898\AA. If the feature at 4881\AA\ is confirmed as \niii\ emission line, so that at 4898\AA\ would also be identified as another \niii\ 
  line (see Table~2).
  {\it SySt--3}:  
  As in SySt--2, \heii$\lambda$4686 line is very intense. The inset shows the \ha\ line, the only one we 
  measured in the red part of the spectrum. The cool star continuum is very prominent, again making the 
  emission-line flux measurements very uncertain. Another conspicuous feature, in all the three blue 
  spectra above, is the \heii$\lambda$4686\ line, whose intense fluxes (see Table~2) imply T$_{eff}$ $>$ 
  270,000~K in the three cases. Also note that CCD gaps were masked in all the spectra (see Section~2). 
  The resulting spectral discontinuities can be seen, for instance, around 7950, 8050 and 8150\AA\ in the 
  upper-right panel.
  }
   \label{fig_spec1}
\end{figure}

The emission-line fluxes were measured with the {\sc IRAF} package {\sc SPLOT}. Errors on the fluxes were calculated taking into account the statistical errors in the measurement of the fluxes, as well as
systematic errors (flux calibrations, background determination and sky subtraction). The resulting errors are given in the fluxes table (Table~2). The Balmer Decrement was used as an estimation of the internal dust attenuation of the SySts. Given that \hd\ and \hg\ were measured at a relatively noisier part of the spectrum, they are not useful for deriving the extinction correction constant, \cbeta\. Thus, \cbeta\ was determined comparing the observed Balmer I (\ha)/I (\hb) ratio with its theoretical value, 2.85 (Osterbrock \& Ferland 2006). The \cbeta=0.445$\pm0.04$ obtained with the Balmer Decrement, for \n205~SySt--1, compares nicely with the mean \cbeta\ we found in G14 for the PNe of \n205, 0.38$\pm$0.09. The \cbeta\ for the other two systems, on the other hand, are very low: 0.168$\pm$0.078 and 0.00$\pm0.54$ for \n205~SySt--2 and 3, 
respectively. Though low, such values are not uncommon within those of the nebular systems either in F1 or in F2 of G14 (see their Table~A1). 

The H$\beta$ flux (F$_{{\rm H}\beta}$), the extinction correction constant (\cbeta), all the other extinction corrected line intensities (I$_{\lambda}$) as well as emission-line fluxes (F$_{\lambda}$), are shown in Table~2.

\section{The Systems' Discovery}

The three emission-line objects were serendipitously discovered in the spectroscopic study of \n205.
Their spectra revealed features that are characteristic of symbiotic systems (see, e.g., Belczy\'nski et al. 2000). In the optical, the spectra of symbiotics are indeed notable due to the absorption features and continuum of late-type M giants, strong nebular emission lines of Balmer \hi, the recombination lines of \hei\ and \heii\ and the forbidden lines of \oii, \neiii, \nev\ and \fevii. In more detail, the well-known criteria to identify symbiotic stars (Belczy\'nski et al. 2000) are the following:  {\em i)} The presence of the absorption 
features of a late-type giant (like titanium  and vanadium oxides, TiO and VO, H$_2$O, CO and CN bands, as well as Ca~{\sc i}, Ca~{\sc ii}, Fe~{\sc i} and Na~{\sc i} absorption lines). {\em ii)}  The presence of strong \hi\ and \hei\ emission lines and either emission lines of ions with an ionization potential of at least 
35~eV (e.g. \oiii), or an A- or F-type continuum with additional shell absorption lines from \hi, \hei, and singly-ionized metals. The latter corresponds to the appearance of a symbiotic star in outburst. {\em iii)} The presence of the $\lambda$6830\AA\ emission feature, even if no feature of the cool star is found. Also, very recently, Lee, Heo \& Lee (2014) identified another Raman--scattered Ne~{\sc vii} $\lambda$973 at 4881\AA\ line, in the spectrum of the Galactic symbiotic star V1016 Cygni. The latter scattering process is again a unique characteristic of symbiotic systems.

The three systems in NGC~205, hereafter \n205\ SySt--1, \n205\ SySt--2, \n205\ SySt--3, show the following characteristics that led us to classify them, respectively, as true, possible and likely symbiotic binaries: 
\begin{itemize}   
\item[1.]
\n205\ SySt-1 shows the \ovi\ Raman scattered lines at 6830\AA, 7088\AA, unique signatures seen only in symbiotic stars (Schmid 1989). These emission lines provide a firm classification criterion of symbiotic systems, even when the red giant continuum is not directly observed, as discussed above. Moreover, in the present case the absorption features of the M giant component are clearly seen. The discovery of \n205\ SySt-1  is similar to that of other SySts in the Local Group (LG) galaxies, as in the case of NGC~6822 (Kniazev et al. 2009) and NGC~185 (Gon\c calves et al. 2012), in which the \ovi\ Raman scattered lines were also detected. 

\item[2.]
\n205\ SySt-2, on the other hand,   
has an extremely intense \heii~4686\AA\ emission line, property that is included  among the classification criteria of SySts (Fig. 2, middle panel). As a consequence of such an intense doubly ionized He emission, the effective temperature (\teff) of the star that ionizes the gas should be tremendously high (see next section; 4.1).  In spite of a high \teff, it is puzzling the fact that higher ionization forbidden lines are not detected in its spectrum, unlike in SySt-1. A possible reason would be the fact that the present system is significantly fainter than the latter (Table~2 shows \hb\ fluxes of 3.70 and 0.74~10$^{-16}$~erg~cm$^{-2}$ ~s$^{-1}$ for SySt-1 and 2, respectively).  In addition, and as far as we know for the first time in an extragalactic SySt, we tentatively identified the Ne~{\sc vii} 
$\lambda$973 at 4881\AA\ Raman-scattered line in the spectrum of SySt-2. This feature can be clearly seen in the inset of the middle panel of Figure 2 (and in Table~2). However, also considering the fact that in the present case only \nevii\ is present, with no sign of the \ovi\ Raman scattered line (both were found at the same epoch in V1016; Lee et al. 2014), the emission line at 4881\AA\ could as well be a N~{\sc iii} line. In this case the line at $\lambda$4898\AA, also present in the spectrum and for which we could not easily find an identification, might be associated with N~{\sc iii} too. Considering the low-resolution of the present spectrum, the other lines from the N~{\sc iii} multiplet are blended with \hb, further complicating a strong identification as N~{\sc iii}. And, a third possible identification for the 4881\AA\ emission-line would be [Fe~{\sc iii}]. Altogether the latter arguments imply that the other possible identifications (N~{\sc iii} and [Fe~{\sc iii}]) are at least equally possible as Ne~{\sc vii} $\lambda$973 at 4881\AA\ Raman-scattered line.

\item[3.]
\n205\ SySt--3 was discovered because of the clear presence of a cool stellar continuum in the red part of 
its spectrum, compatible with that of a M giant. This symbiotic star, as well as the one discovered in the LG dwarf irregular galaxy IC10 (Gon\c calves et al. 2008), though not showing the Raman scattered lines, shares some other properties of Galactic sources. The latter includes the moderate to high-ionization [Fe~{\sc v}] line, at 4181\AA\ (see, for instance, Gutierrez-Moreno \& Moreno 1996). These two conspicuous properties of SySts lead us to conclude that SySt-3 is, likely, a symbiotic system.
\end{itemize}

We compare our detections with other literature surveys in \n205\ also interested in emission-line populations, in particular PNe:  Ford et al. (1973), Ciardullo et al. (1989), Richer \& McCall (1995), 
Corradi et al. (2005), Merrett et al. (2006) and Richer \& McCall (2008). All these surveys studied this galaxy using narrow-band imaging and/or optical spectroscopy. From these works, only the spectroscopic survey of Richer \& McCall (1995 and 2008) could had found out our symbiotics. However, targets in Table~1 are not included among theirs. The \n205\ SySt--1 is also not listed among the objects analysed-- via narrow-band imaging-- by Corradi et al. (2005), while they detected the other two systems and the correspondent ID in their survey are also given in Table~1. The most probable reason why SySt--1 was not detected by Corradi et al. (2005) is  that the completeness limit of their survey was only 50\% in mag(\oiii)=~24.5 -- where mag(\oiii)=-2.5 $\log$ F(\oiii5007) - 13.74 (Ciardullo et al. 1989). \oiii\ magnitudes of 24.8, 24.4 and 23.6 are the present magnitudes of SySt--1, 2 and 3, respectively.

In Figure~1 we show the fields we observed with GMOS, and the finding charts of the three newly discovered 
symbiotic stars. Their coordinates, as well as their IDs in G14 and Corradi et al. (2005) are given in Table~1. Compositions of the blue plus red portions of the GMOS spectra of NGC~205 symbiotics are shown in Figure~2, where the main properties that make of them symbiotic systems are highlighted. 

\section{Characterization of the Symbiotic Systems}
\subsection{The Ionised Nebulae and the Ionising  Sources}

We have determined the above discussed extinction coefficient, and corrected all the measured fluxes for further analysis, despite the fact that the Balmer line ratios in many symbiotic nebulae indicate self-absorption effects (due to high densities). In those circumstances, the standard methods to estimate reddening would not apply (possibly the case in SySt--2 and 3). Though most of the line ratios that would provide electron temperatures and densities comes to be out of the sensitive range (see Osterbrock \& Ferland 2006), the sulphur doublet \sii~$\lambda\lambda$6716, 6731 allowed us to estimate an electron density of 6,000~cm$^{-3}$ for SySt--1 (since the \oiii-temperature sensitive ratio is below the lower limit, for the above calculation we adopted \te=21,000~K). The density sensitive lines were not measured in the remaining cases. \oiii\ electron temperatures can be derived from the \oiii~$\lambda\lambda$4959,5007 doublet to the $\lambda$4363 line ratio. Relatively high \te\oiii\ were found for SySt--2 and 3, of $>$21,000~K and 14,500~K, respectively. Also, using the extinction-corrected integrated flux of \heii~$\lambda$4686 (Kaler \& Jacoby 1989), we derived T$_{eff}$ of $>$340,000, $>$340,000 and 270,000~K as the temperature of the hot companion, which represents the main nebular ionising source, of SySt--1, 2 and 3, respectively.

The above derived diagnostics constitute further evidences in favour of our targets being representative of  symbiotic systems, since \te\ and T$_{eff}$  are much higher than the typical values for these parameters found either in PNe or compact \hii\ regions.
 
\subsection{The Spectral Classification of the Cool Companions}

\begin{figure} 
  \centering
  \includegraphics[width=8.5truecm]{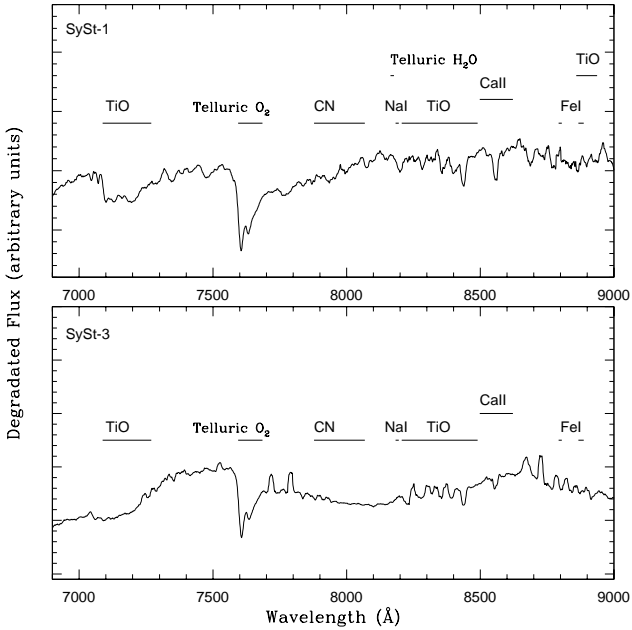} 
  \caption{The 6900 to 9000\AA\ range of the red spectrum of \n205~SySt--1 and 3, smoothed to match 
  the spectral resolution of the Kirkpatrick et al. (1991) catalogue (18\AA\ per pixel). Following 
  Kirkpatrick et al. (1991), the most prominent features, key to determine the spectral type, are 
  identified. i) The CN bands are obvious in supergiants, weaker in giants and not seen in dwarfs. 
  ii) The TiO whose head band is at 8206\AA\ is very strong in mid-M. iii) The CaII bands are strong 
  from late-K to mid-M, strongest in supergiants and very weak in dwarfs. iv) Fe~I absorptions (at 8793 
  to 8805$~\AA$) are found in giants as well as supergiants.  
   }
\end{figure}

\begin{table}
\centering
\begin{minipage}{85mm}
{\tiny  
\caption{Observed fluxes (F$_{\lambda}$) and extinction corrected intensities (I$_{\lambda}$) of 
\n205 SySt--1, 2 and 3. The observed \hb\ flux is in units of 10$^{-16}$~erg~cm$^{-2}$ ~s$^{-1}$, and both I$_{\lambda}$ and F$_{\lambda}$ are normalized to \hb=100. Errors, obtained for different flux ranges,  
are: up to 6, 15, 25, 40, $>$50\% for F$_{\lambda}$ $>$ 100, 10-100, 1-10, 0.1-1.0, and 0.01-0.1~F$_{{\rm H}\beta}$, respectively. ``:'' indicate very uncertain measurements, due to loosely defined continuum.} 
\begin{tabular}{@{}cclcrl@{}}
\hline
F$_{{\rm H}\beta}$ & \cbeta\         & Ion & $\lambda$ (\AA) & I$_{\lambda}$ & F$_{\lambda}$ \\  
\hline
\multicolumn{6}{c}{\bf \n205~SySt--1} \\ 
\hline
3.700 & 0.445$\pm$0.04	 & [Ne~{\sc iii}]     & 3868   & 16.60  & 12.30   \\ 
      &          & [Ne~{\sc iii}]+He~{\sc i}  & 3968   & 22.20  & 16.87   \\ 
      &          & N~{\sc iii}                &  4003  & 19.66  & 15.08   \\ 	 
      &          &  -                       &  4084  & 10.27  & 8.064   \\ 	 
      &          & H$\delta$	              &  4100  & 36.41  & 28.73   \\ 
      &          & H$\gamma$	              &  4340  & 72.00  & 61.22   \\ 
      &          & [O~{\sc iii}]              &  4363  & 42.92  & 36.77   \\ 
      &          &He~{\sc i}	              &  4437  & 9.926  & 8.706   \\ 		 
      &          &C~{\sc ii}+O~{\sc ii}       &  4491  & 8.156  & 7.280   \\ 
      &          &Fe~{\sc ii} 		      &  4584  & 5.230  & 4.810   \\	 
      &          &	O~{\sc ii}	      &  4638  & 5.040  & 4.713   \\	 
      &          & He~{\sc ii}                &  4686  & 124.2  & 117.8   \\ 	 
      &          & H$\beta$	              &  4861  & 100.0  & 100.0   \\ 	 
      &          & [O~{\sc iii}]              &  4959  & 34.19  & 35.06   \\ 		 
      &          & [O~{\sc iii}]              &  5007  & 102.2  & 106.1   \\ 
      &          & He~{\sc i}		      &  5016  & 4.032  & 4.200   \\ 
      &          & [Fe~{\sc vii}]             &  5159  & 7.891  & 8.500   \\ 			 
      &          & [Fe~{\sc ii}]	      &  5261  & 3.112  & 3.429   \\ 	 
      &          & Fe~{\sc iii}               &  5275  & 3.756  & 4.151   \\ 
      &          &	[Ca~{\sc v}]	      &  5308  & 3.113  & 3.493   \\ 	 
      &          & [Fe~{\sc ii}]              &  5376  & 1.384  & 1.562   \\ 	 
      &          & He~{\sc ii}                &  5412  & 4.917  & 5.587   \\ 
      &          & [Fe~{\sc ii }]             &  5496  & 2.219  & 2.562   \\ 				 
      &          & [Fe~{\sc ii}]              &  5582  & 21.25  & 24.93   \\ 		 
      &          &	V~{\sc ii}            &  5618  & 4.195  & 4.950   \\ 		  
      &          &	[Fe~{\sc vii}]        &  5720  & 10.19  & 12.23   \\ 	  
      &          & [N~{\sc ii}]               &  5755  & 2.189  & 2.648   \\ 
      &          & He~{\sc i}		      &5876 & 11.12 & 13.70 \\
      &          & [Ca~{\sc v}]+[Fe~{\sc ii}] &  6086  & 12.84  & 16.29   \\  
      &          & [O~{\sc i}]                &  6300  & 3.534  & 4.621   \\ 
      &          &  [S~{\sc iii}]             &  6312  & 1.414  & 1.853   \\     
      &          & H$\alpha$	              &  6563  & 307.2  & 416.4   \\ 
      &          & [N~{\sc ii}]               &  6584  & 2.774  & 3.770   \\ 
      &          & He~{\sc i}                 &  6678  & 2.253  & 3.101 :    \\ 
      &          & [S~{\sc ii}]               &  6717  & 0.508  & 0.703 :   \\ 		 
      &          & [S~{\sc ii}]               &  6731  & 0.865  & 1.199 :   \\  
      &          & [Fe~{\sc vi}]              &  6739  & 0.610  & 0.847 :   \\						
      &          & O~{\sc vi} Raman	      &  6830  & 28.41  & 39.93 :   \\
      &          & He~{\sc i}	              &  7065  & 2.841  & 4.123 :   \\ 
      &          & O~{\sc vi} Raman      &  7088  & 5.426  & 7.899 :   \\ 		    
      &          & [O~{\sc ii}]               &  7320. & 0.890  & 1.338 :   \\  
      &          & [O~{\sc ii}]               &  7330. & 0.982  & 1.478 :   \\
      &          & [O~{\sc i}]                &  8447. & 3.776  & 6.550 :   \\ 
\hline
\multicolumn{6}{c}{\bf \n205~SySt--2} \\ 
\hline
0.740&0.168$\pm$0.078& H$\gamma$     & 4340  & 80.42 & 75.69   \\
     &        & [O~{\sc iii}]        & 4363  & 52.24 & 49.31   \\ 
     &      	  & He~{\sc ii}      & 4686  & 137.4 & 134.8   \\ 
     &      	  & H$\beta$	     & 4861  & 100.0 & 100.0   \\ 
     &      	  & Ne~{\sc vii} Raman~?, N~{\sc iii}~?, [Fe~{\sc iii}]~? & 4881  & 30.11 & 30.16   \\ 
     &      	  & N~{\sc iii}?               & 4898  & 13.37 & 13.42   \\
     &      	  & [O~{\sc iii}]    & 4959  & 233.5 & 235.8   \\ 
     &      	  & [O~{\sc iii}]    & 5007  & 694.3 & 704.5   \\ 
     &      	  & He~{\sc i}       & 5876  & 12.44 & 13.45   \\
     &      	  & H$\alpha$	     & 6563  & 293.9 & 329.7 : \\ 
     &      	  &  -                & 6783  & 34.00 & 38.58 :\\
     &      	  &  -                & 7167  & 11.06 & 12.80 :\\
     &      	  &  -                & 8107  & 19.82 & 24.05 :\\
     &      	  &  -                & 8782  & 40.42 & 50.41 :\\ 
     &      	  & [S~{\sc iii}]    & 9069  & 24.79 & 31.19 :\\      
\hline
\multicolumn{6}{c}{\bf \n205~SySt--3} \\ 
\hline
1.263& 0.00$\pm$0.54&    [O~{\sc ii}]  & 3729  &   21.6  &   21.6 \\ 
     &      	  &  [Ne~{\sc iii}]    & 3868  &   32.7  &   32.7 \\ 
     &      	  &  H$\delta$	       & 4100  &   25.3  &   25.3 \\ 
     &      	  &  [Fe~{\sc v}]     & 4180  & 35.3  &   35.3 \\ 
     &      	  &  H$\gamma$	       & 4340  &   54.8  &   54.8 \\
     &      	  &  [O~{\sc iii}]     & 4363  &   16.4  &   16.4 \\ 
     &            &   -                & 4553  &   82.89 &   82.89 \\
     &      	  &  He~{\sc ii}       & 4686  &   84.8  &   84.8 \\ 
     &      	  &  H$\beta$	       & 4861  &  100.0  &  100.0 \\ 
     &      	  &  [O~{\sc iii}]     & 4959  &  295.4  &  295.4 \\ 
     &      	  &  [O~{\sc iii}]     & 5007  &  860.1  &  860.1 \\ 
     &      	  &  -                 & 5343  &   22.1  &   22.1 \\ 
     &      	  &  [Fe~{\sc ii}]     & 5582  & 220.5  &  220.5 \\ 
     &      	  &  -                 & 5586  & 76.0   &   76.0 \\  
     &      	  &  H$\alpha$	       & 6563  &  179.2  &  179.2 : \\ 
\hline
 \end{tabular}
}
\end{minipage}
\label{tabPN_flux}
\end{table}

Having only the optical spectrum of the system it is hard to obtain the properties of the cool companion. However, limits to the spectral types based on the red part of the optical spectrum of symbiotic stars are possible (Kenyon \& Fern\'andez-Castro 1987). The latter authors, using their own data as well as the literature, pointed out that:

\begin{itemize}
\item there is no evidence for measurable TiO absorption in stars earlier than the spectral type K4;
\item TiO indices increase monotonically with spectral type from K4 to M6;
\item VO bands, like the one at 7865\AA, appear only in giants cooler than $~$M5;
\item the amount of VO absorption is negligible for stars earlier than M4, but rises rapidly for M4-M7.
\end{itemize}

\noindent Taking all these limits into account, and the spectra we show in Figure~3 (on which neither the 7865\AA, nor any other VO features are present), it is clear that the stars there represented have spectral types from K4 to M4. In order to avoid the tricky task of defining a common continuum, below which the TiO absorption indices could be measured (Kenyon \& Fern\'andez-Castro 1987), we refer the reader to Kirkpatrick et al. (1991), who provide an extensive spectral catalogue for late-type stars from classes K5-M9, covering the wavelengths from 6900 to 9000\AA. This region encompasses TiO and VO bands --key bands to classify early and late M stars. Table~5 and Figure~5 of their paper give the features found in late-K to late-M red spectra, and provide us with a more accurate range of spectral classes for the cool stars of NGC~205 SySt--1 and SySt--3, as being K5 to M2 red giant (see Kirkpatrick's Fig.~5a and 5b against our Figure~3). To perform this classification, our GMOS spectra had their spectral resolution smoothed to ~18 \AA\ per pixel, in order to better compare to  Kirkpatrick et al. (1991) catalogue. To close this analysis, we remember that the \teff\ of K and M stars ranges, roughly, from 3,700 to 5,200~K and from 2,400 to 3,700~K, respectively.

\subsection{The Spectral Energy Distribution for SySt-1}

\begin{figure} 
  \centering
  \includegraphics[width=8.5truecm]{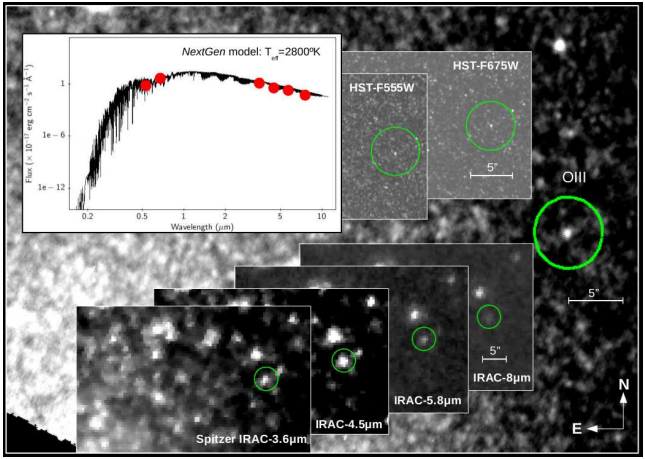} 
  \caption{SED-fitting and multi-wavelength observations, from HST and Spitzer-IRAC, are superimposed 
  over the \oiii\ image of SySt--1. SED points are photometrically extracted from the observations and 
  the error bars are smaller than the dot size. The best SED-fitting, carried out with VOSpec tool, shows 
  a NextGen Model (Hauschildt et al. 1999) with T=2,800~K. 
  }
\end{figure}

Due to the extragalactic location of the stars, resolved photometric information is only available from 
selected telescope-instrument combinations, producing small PSFs over pixels of an appropriate scale. Neither optical nor IR counterparts had been found for SySt-2 and SySt-3, probably because they are significantly much fainter than SySt-1 (the F$_{{\rm H}\beta}$ of these two sources are $\sim$7 and $\sim$3 times lower than in SySt-1).  

The Spectral Energy Distribution (SED) of SySt-1 has been constructed with available information from the optical and IR spectral ranges. In addition to our optical \oiii\ image of SySt-1, presented in Section~2, the object has been unambiguously detected by HST-WFPC2 and Spitzer-IRAC. All images have been retrieved from public archives, those from the HST come from the Mikulski Archive for Space Telescopes (MAST), while those from Spitzer-IRAC have been obtained from the NASA/IPAC Infrared Science Archive (IRSA). The HST-WFPC2 images, with detections in four filters (F255W, F336W, F555W and F675W) have a 0.0996\arcsec /pixel scale, with a PSF of typically 2 pixels FWHM. Aperture photometry was carried out by ourselves 
using the {\sc IRAF} package {\sc APPHOT} and converted into flux density. The IR images from Spitzer-IRAC provide object detections in the four IRAC (Infrared Array Camera) wavelengths (3.6, 4.5, 5.8 and 8$\mu$m). The Spitzer-IRAC pixel size is 1\farcs2 /pixel and the typical FWHM of the IRAC-1 PSF is 2.7 pixels. The IR photometric values for the flux density are taken directly from the Spitzer-IRAC pipeline. We have carefully checked that the lack of detection of the SySt-1 in several popular surveys as, SDSS, 2MASS, WISE is simply due to insufficient spatial resolution.

The final SED has been constructed with six points from both the HST-WFPC2 (2) and Spitzer-IRAC (4). The SED points corresponding to filters HST-F255W and F336W have been excluded. Although visually discernible, their photometric signal to noise is too low ($\sim$ 2). Flux values have been corrected for Galactic extinction, which according to Schlafly \& Finkbeiner (2011) amounts to A$_V$=0.170 mag. In addition to Galactic extinction, internal dust attenuation in the SySt-1 has an extinction constant \cbeta=0.45 (see Section 2.1), calculated via the Balmer Decrement. Assuming a R$_V$=3.1 extinction law, we deduce an 
A$_V$=0.955 mag, also used to correct the observed SED values for reddening. 

Model fitting has been carried out with the VOSpec tool from the ESAC Virtual Observatory Project. In addition to the crude Black Body fitting, which gives T$_{eff}$ = 2,455~K, we have fitted cool-star NextGen Models (Hauschildt et al. 1999), which produce best fitting for T$_{eff}$ = 2,800~K. We must emphasize that --though the agreement of the latter effective temperatures with the M red giant classification we obtained in Section~3.2 is very good-- fittings are just indicative, because the HST optical SED points are probably overestimations of the continuum, as they include intense emission lines in their passbands, \oiii5007 in the F555W and H$\alpha$ in the F675W.

\subsection{Diagnostic Diagram: Separating SySts from PNe}

\begin{figure} 
  \centering
  \includegraphics[width=8.5truecm]{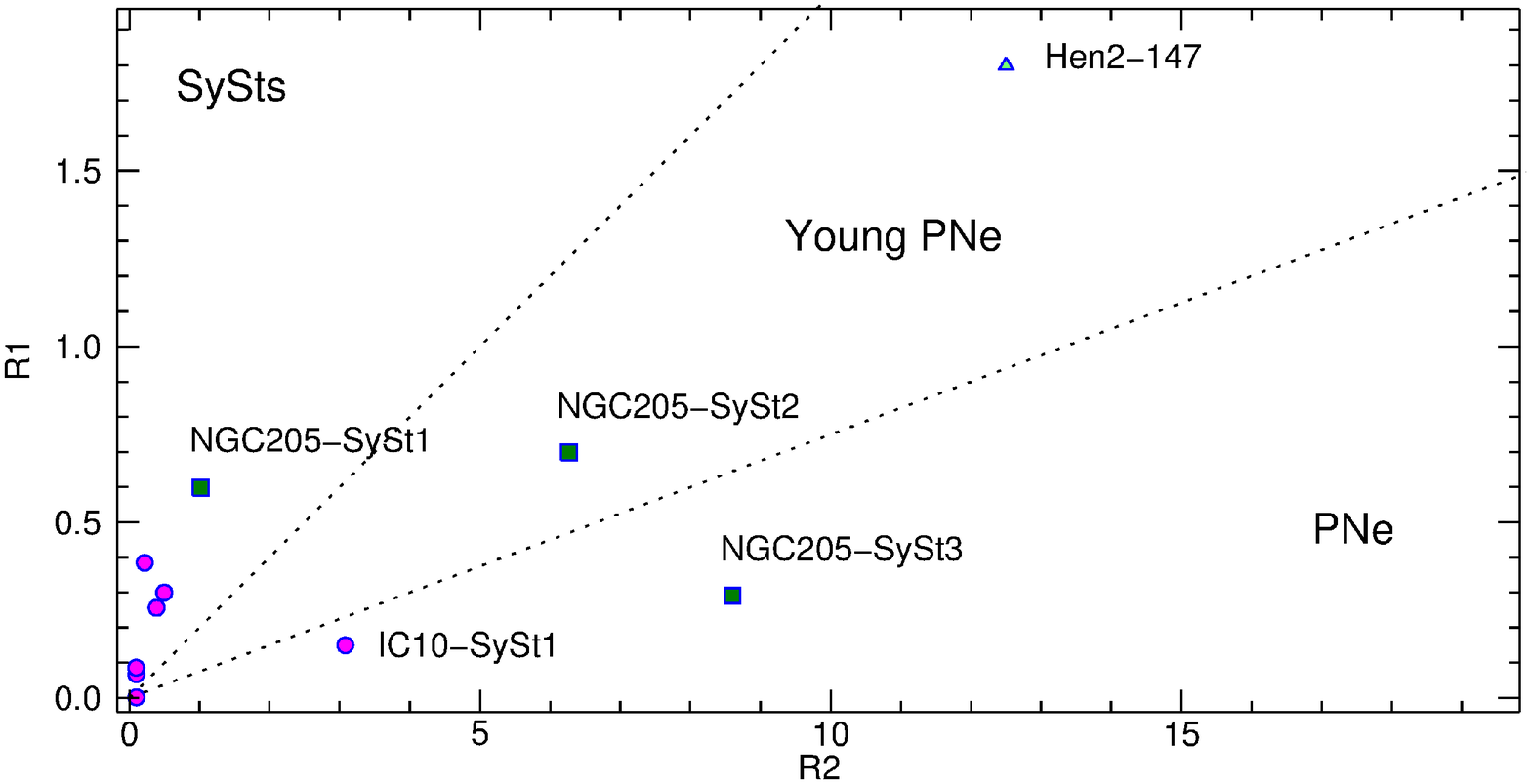} 
  \caption{The diagnostic diagram to separate SySts from PNe: R1=\oiii4363/\hb\ vs. 
  R2=\oiii5007/\hd; as proposed by Gutierrez-Moreno et al. (1995). Boxes represent the objects discussed 
  in the present paper, whose line ratios are presented in Table~2. The circles represent 7 symbiotic 
  systems found in LG dwarf galaxies, for which the line ratios were published (LMC1, Morgan (1992); 
  LMC-Sanduleak and LMC-S63, Allen 1980); or for which limits can be obtained from the published 
  spectra (IC10, Gon\c calves et al. 2008; NGC~185, Gon\c calves et al. 2012; NGC~6822, Kniazev et al. 
  2009). The triangle represents the Galactic SySt Hen~2-147 whole line ratios were published by 
  Mikolajewska et al. (1997).}
\end{figure}

Gutierrez-Moreno et al. (1995) demonstrated that SySt can be separated from PNe in \oiii4363/\hg\ vs. \oiii5007/\hb\ diagram, being this separation mainly due to difference in physical conditions of the 
two kind of objects, especially the significantly lower density in PNe. The same is true for all the 35 recently discovered SySts in M31 (see Fig. 5 in Mikolajewska et al., 2014). 

In Figure~5 we place in the Gutierrez-Moreno et al. (1995) diagram all the SySt of the LG dwarf galaxies, for which the emission-line ratios are available. Most of them are indeed located in the loci of the SySts as studied by Gutierrez-Moreno et al. (1995) in the Galaxy, and by Mikolajewska et al. (2014) in M31. From this figure it is clear that our SySt-1 is a true symbiotic system, and, at a first glance, the other two objects would be discarded as such. 

However it is important to note that another symbiotic discovered in IC10, a LG dwarf irregular galaxy (Gon\c calves et al. 2008), is also misplaced in this diagram. In the case of the IC10 SySt-1, alike SySt-2 and 3 of NGC~205, neither highly ionized emission-lines nor Raman scattered lines are found in the spectrum. The basis for the classification of the IC10 SySt-1 was the extreme similarity of its red spectrum with that of a well-known Galactic symbiotic,  Hen~2-147 (Munari and Zwitter 2002), on which the cool companion features are clearly present. We also searched in the literature the emission-line ratios of Hen~2-147 to properly place it in the Gutierrez-Moreno et al. (1995) diagnostic diagram. Unfortunately, these ratios were not available in the literature. Thus, we used the Munari and Zwitter (2002) online spectrum of the system to roughly estimate the ratios. The result (\oiii4363/\hg~1.5 and \oiii5007/\hb~12-15) again moves this Galactic SySt in the loci of PNe, more exactly in the regime of young PNe. Because of their high densities, young PNe and SySts occupy the same region in this diagram, as pointed out by Gutierrez-Moreno et al. (1995) and Pereira \&  Miranda (2005). Therefore, what we show in Fig.~5 is that our SySt-2 as well as the known Galactic SySt Hen~2-147 occupy the young PN region of the diagram, whereas our SySt-3 and the known SySt-1 of IC10 are placed in the region of the evolved PNe. Altogether this seems to indicate that not only 2 of the (possible and likely)  SySts in NGC~205,  but also other known SySts are misplaced in this diagram, probably due to other effects than the density. Gutierrez-Moreno et al. (1995) also pointed out the fact that very young PNe and D-Type SySts are sometimes impossible to be separated in their diagram.

Noting that the Milky Way has similar metallicity as M31 (Mikolajewska et al. 2014), and that most of the SySts of these two spiral galaxies are placed in similar regions of the Gutierrez-Moreno et al. (1995) diagnostic diagram, we also investigate the possible metallicity effects in the diagram. We did so by placing on it the SySts of the lower-metallicity LG dwarf galaxies. Fig. 5 shows that, even at the lower metallicities of the LG dwarf galaxies, this effect is insufficient to significantly change the loci of PNe and SySts in the diagram. 

	Finally, it is worth mentioning the fact that, at variance with most of the SySt candidates that were identified as so via H$\alpha$ narrow-band imaging --and then their spectra were taken to confirm the classification of the objects-- in the present paper the narrow-band images were centred on the \oiii~5007 filter. This is probably why our SySts are much brighter in \oiii\ than the majority of the known SySts.

\section{Summary}
In this letter, we report the discovery of the first three true, likely and possible SySts of the dwarf spheroidal galaxy NGC~205. Photometric and spectroscopic data of these objects were obtained with the Gemini Multi--Object Spectrograph (GMOS), installed on the Gemini North 8.1--m telescope. Among the detection of several \oiii\ line emitters in the NGC~205, three objects were identified as SySts.

The classification of these sources as SySts was done based on the identification of the Raman scattering 
 O~{\sc vi} lines around $\lambda\lambda$6830,7088\AA\ and Ne~{\sc vii} of 4881\AA\ (SySt-1 \& SySt-2), 
 the direct observation of the red continuum of the cool companion (SySt-1 \& SySt-3), the extremely high intensity of the He II 4846\AA\ recombination line (SySt-1, SySt-2 \& SySt-3), and the presence of moderate to high-ionization forbidden lines (for instance \fev\ and \fevii), in SySt-1 and 3. If confirmed as a true symbiotic binary, SySt-2 will be the second known symbiotic star (and the first extragalactic one) that shows the Ne~{\sc vii} Raman scattering line at 4881\AA.

The effective temperature of the hot companions were estimated $>$340,000~K for SySt-1 and SySt-2 and $\sim$270,000~K for SySt-3. These extremely high temperatures are consistent with the detection of high excitation emission lines in the spectra of SySt-1 and 3 (e.g  O~{\sc vi}, [Fe~{\sc vii}], Ne~{\sc vii}).  Being the faintest of the three systems here studied, SySt-2 is an exception in having \oiii\ as its highest ionization forbidden line.

By examining the red continuum spectra of the cool companions, they were classified as K5 to M2 red giants.  It is worth mentioning that the classification of the cool companion in SySt-1 as a M2 III is consistent with a cool--star NextGen stellar atmospheric model of T$_{eff}$ = 2,800~K. This kind of analysis as well as the classification of the cool companion is, unfortunately, not feasible for SySt-2.

We  discuss the location of the newly discovered systems in the Gutierrez-Moreno et al. (1995) diagram, meant to separate SySts from PNe in \oiii4363/\hg\ vs. \oiii5007/\hb\ plane. From their location in this diagnostic diagram, only  SySt-1 would be a true symbiotic system. 
Our SySt-2 as well as the known Galactic SySt Hen~2-147 occupy the young PN region of this diagram, whereas our SySt-3 and the known SySt-1 of IC10 are placed in the region of the evolved PNe. Altogether this seems to indicate that not only 2 of the (possible and likely)  SySts in NGC~205 but also other known SySts are misplaced in this diagram, probably because of the interplay of other effects than the physical conditions of the two kinds of nebulae.  
Gutierrez-Moreno et al. (1995) also pointed out the fact that very young PNe and D-Type SySts are sometimes impossible to be separated in their diagram. We finally note that though the different metallicities of the 
spirals (Milky Way and M31) versus the dwarf galaxies of the LG might play a role in the diagram, the use of available data indicates that this effect does not appear to be strong enough to significantly change the location of SySts and PNe in the Gutierrez-Moreno et al. (1995) diagram.

\section{Acknowledgments}
Authors are very grateful to Joanna Mikolajewska, the referee, for her criticisms to the manuscript that allow us to correct mistakes of the first version and add the important discussios to the paper. We also thanks Claudio P. Bastos and Roberto Costa for their critical reading of the manuscript and a number of fruitful discussions. DRG kindly acknowledges the Instituto de Astrof\'isica de Canarias (IAC), for their hospitality, where part of this work was done. LM acknowledges financial support from PRIN MIUR 2010-2011, project \lq\lq The Chemical and Dynamical Evolution of the Milky Way and Local Group Galaxies'', prot. 2010LY5N2T. SA is supported by CAPES  post-doctoral scholarship under the program \lq\lq Youth Talents", A035/2013. This work was partially supported by FAPERJ's grant E-26/111.817/2012 and CAPES's grant A035/2013.

{}

\label{lastpage}


\begin{thebibliography}{}
\bibitem[\protect\citeauthoryear{Allen}{1980}]{}
Allen D. A., 1980, ApL, 20, 131

\bibitem[\protect\citeauthoryear{Belczy\'nski et al.}{2000}]{}
Belczy\'nski K., Mikolajewska J., Munari U., Ivison R. J., \& FriedjungX M., 
2000, A\&AS, 146, 407

\bibitem[\protect\citeauthoryear{Ciardullo et al.}{1989}]{} 
Ciardullo R., Jacoby G. H., Ford
H., \& Neill J.D., 1989, ApJ, 339, 53

\bibitem[\protect\citeauthoryear{Corradi et al.}{2005}]{} 
Corradi R.~L.~M., et al., 2005, A\&A, 431, 555 

\bibitem[\protect\citeauthoryear{Corradi \& Magrini}{2006}]{}
Corradi R.~L.~M., \& Magrini L., 2006, in ``Planetary nebulae beyond the Milky Way", 
Proceedings of the ESO workshop, Springer, 2006, p.36

\bibitem[\protect\citeauthoryear{Ford, Jacoby \& Jenner}{1973}]{}
Ford H. C., Jenner D. C., \& Epps, Harland W., 1973, ApJ, 183, L73

\bibitem[\protect\citeauthoryear{Gon{\c c}alves et al.}{2008}]{} 	
Gon\c calves D. R., Magrini L., Munari U., Corradi R. L. M., \& Costa R. D. D.,
2008, MNRAS, 391, L84
	
\bibitem[\protect\citeauthoryear{Gon{\c c}alves et al.}{2012}]{} 
Gon{\c c}alves D.~R., Magrini L., Martins 
L.~P., Teodorescu A.~M., Quireza C., 2012, MNRAS, 419, 854 

\bibitem[\protect\citeauthoryear{Gon{\c c}alves et al.}{2014}]{} 
Gon{\c c}alves D.~R., Magrini L., Teodorescu A.~M., \& Carneiro C. M., 2014, MNRAS, 444, 1705

\bibitem[\protect\citeauthoryear{Gutierrez-Moreno et al.}{1995}]{}
Gutierrez-Moreno A., Moreno H., \& Corte G., 1995, PASP, 107, 462

\bibitem[\protect\citeauthoryear{Gutierrez-Moreno \& Moreno}{1996}]{}
Gutierrez-Moreno A., \& Moreno H., 1996, PASP, 108, 972

\bibitem[\protect\citeauthoryear{Hauschildt et al.}{1999}]{}
Hauschildt P. H., Allard F., Baron E., \& Schweitzer A., 1999, ApJ, 312, 377 

\bibitem[\protect\citeauthoryear{Kaler \& Jacoby}{1989}]{}
Kaler J. B., Jacoby G. H., 1989, ApJ, 345, 871

\bibitem[\protect\citeauthoryear{Kenyon \& Fernandez-Castro}{1987}]{}
Kenyon S. J., \& Fernandez-Castro T., 1987, AJ, 93, 938

\bibitem[\protect\citeauthoryear{Kirkpatrick et al.}{1991}]{}
Kirkpatrick J. D., Henry T. J., \& McCarthy D. W. Jr., 1991, ApJSS, 77, 419

\bibitem[\protect\citeauthoryear{Kniazev et al.}{2009}]{}
Kniazev A. Y., et al., 2009, MNRAS, 395, 1121

\bibitem[\protect\citeauthoryear{Lee et al.}{2014}]{}
Lee H-W, Heo J-E, \& Lee B-C,  2014, MNRAS, 442, 1956

\bibitem[\protect\citeauthoryear{Leedj\"arv et al.}{2004}]{}
Leedj\"arv L., Burmeister M., Miko\'lajewski M., Puss A., Annuk K., 
Ga\'lan C., 2004, A\&A, 415, 273

\bibitem[\protect\citeauthoryear{Magrini et al.}{2003}]{}
Magrini L., Corradi R.~L.~M., \& Munari U., 2003, ASPC, 303, 539

\bibitem[\protect\citeauthoryear{Mikolajewska et al.}{2014}]{} 
Mikolajewska J, Caldwell N., \& Shara M. M., 2014, MNRAS, 444, 586

\bibitem[\protect\citeauthoryear{Morgan}{1992}]{}
Morgan D. H., 1992, MNRAS, 258, 639

\bibitem[\protect\citeauthoryear{Magrini, Corradi, 
\& Munari}{2003}]{} Magrini L., Corradi R.~L.~M., Munari U., 2003, ASPC, 303, 539 

\bibitem[\protect\citeauthoryear{Massey et al.}{1988}]{}
Massey P., Strobel K., Barnes J. V., Anderson E., 1988, ApJ, 328, 315

\bibitem[\protect\citeauthoryear{Massey \& Gronwal}{1990}]{}
Massey P., Gronwall C., 1990, ApJ, 358, 344

\bibitem[\protect\citeauthoryear{Mathis}{1990}]{}
Mathis J. S., 1990, ARA\&A, 28, 37

\bibitem[\protect\citeauthoryear{McConnachie}{2012}]{}
McConnachie A. W., 2012, AJ, 144, 4

\bibitem[\protect\citeauthoryear{Merrett et al.}{2006}]{}
Merrett H. R., et al., 2006, MNRAS, 369, 120

\bibitem[\protect\citeauthoryear{Mikolajewska et al.}{1997}]{}
Mikolajewska J., Acker A., \& Stenholm B.,  1997, A\&A, 327, 191

\bibitem[\protect\citeauthoryear{Mikolajewska et al.}{2014}]{}
Mikolajewska J., Caldwell N., \& Shara M. M., 2014, MNRAS, 444, 586

\bibitem[\protect\citeauthoryear{Munari \& Zwitter}{2002}]{}
Munari U., \& Zwitter T., 2002, A\&A, 383, 188

\bibitem[\protect\citeauthoryear{Osterbrock \& Ferland}{2006}]{}
Osterbrock D. E. \& Ferland G. J., in Astrophysics of gaseous nebulae and active galactic nuclei, 2nd. ed.
 Sausalito, CA: University Science Books, 2006

\bibitem[\protect\citeauthoryear{Pereira \& Miranda}{2005}]{}
Pereira C. B., \& Miranda L. F., 2005, A\&A, 433, 579

\bibitem[\protect\citeauthoryear{Richer \& McCall}{1995}]{}
Richer M. G., McCall M. L., 1995, ApJ, 445, 642

\bibitem[\protect\citeauthoryear{Richer \& McCall}{2008}]{}
Richer M. G., McCall M. L.,  2008, ApJ, 684,1190

\bibitem[\protect\citeauthoryear{Schlafly \& Finkbeiner}{2011}]{}
Schlafly E. F., \& Finkbeiner D. P., 2011, ApJ, 737, 103

\bibitem[\protect\citeauthoryear{Schmid}{1989}]{}
Schmid H. M., 1989, A\&A, 211, L31

\bibitem[\protect\citeauthoryear{Schmid et al.}{1999}]{}
Schmid H. M., et al., 1999, A\&A, 348, 950

\bibitem[\protect\citeauthoryear{Tomova \& Tomov}{1999}]{}
Tomova M. T., Tomov N. A., 1999, A\&A, 347, 151
\end{thebibliography}
\end{document}